\documentclass[aps,prd,onecolumn,superscriptaddress,nofootinbib]{revtex4}
\usepackage{graphicx,amsfonts,amsthm}
\usepackage{amsmath,amssymb}
\usepackage[normalem]{ulem}
\usepackage{color}
\usepackage{tikz} 
\usepackage[pdfborder={0 0 0}]{hyperref}
\usepackage[utf8]{inputenc}
\hypersetup{
	colorlinks=true,
	linkcolor=black,
	citecolor=black,
	filecolor=black,
	urlcolor=black,
}

\definecolor{darkgreen}{rgb}{0.0,0.5,0.0}
\newcommand{\be}{\begin{equation}}
\newcommand{\ee}{\end{equation}}
\newcommand{\bea}{\begin{eqnarray}}
\newcommand{\eea}{\end{eqnarray}}
\newcommand{\dst}{\displaystyle}
\newcommand{\fr}[2]{\frac{{\dst #1}}{{\dst #2}}}
\renewcommand{\Re}{\mathrm{Re }}
\renewcommand{\Im}{\mathrm{Im }}
\newcommand{\doublet}[2]{ \left( \begin{array}{c}#1 \\ #2 \end{array}\right) }

\newcommand{\lr}[1]{ \langle #1 \rangle}

\newcommand{\Z}{\mathbb{Z}}

\newcommand{\toCP}{\xrightarrow{CP}}

\newcommand{\lrpartial}{\,\partial^{\hspace{-7pt}\raise3pt\hbox{\small $\leftrightarrow$}}\!}

\newcommand{\varv}{\mathrm{v}}

\def\GeV{\,{\rm GeV}}

\def\cm{\,{\rm cm}}
\def\s{\,{\rm s}}
\def\SM{\,{\rm SM}}
\def\Yeq{Y_{\rm eq}}
\def\bphi{\bar{\varphi}}

\def\Yt{Y_{t}}
\def\dY{\Delta}
\def\svSM{\lr{\sigma v}_{\!\SM}} 
\def\svSMsym{\lr{\sigma v}_{\!\SM, 0}} 
\def\svC{\lr{\sigma v}_{\rm conv}} 
\def\sC{\sigma_{\rm conv}} 
\newcommand{\derivX}[1]{\dot{#1}} 

\def\lsim{\mathrel{\rlap{\lower4pt\hbox{\hskip1pt$\sim$}}
    \raise1pt\hbox{`$<$}}}         
\def\gsim{\mathrel{\rlap{\lower4pt\hbox{\hskip1pt$\sim$}}
    \raise1pt\hbox{$>$}}}         

\begin{document}
\title{
{\normalsize \hfill CFTP/18-017} \\*[7mm]
Dark matter from $CP$ symmetry of order 4: evolution in the asymmetric regime}

\author{Igor P. Ivanov}\thanks{E-mail: igor.ivanov@tecnico.ulisboa.pt}
\affiliation{CFTP, Instituto Superior T\'{e}cnico, Universidade de Lisboa,
Avenida Rovisco Pais 1, 1049 Lisboa, Portugal}
\author{Maxim Laletin}\thanks{E-mail: maxim.laletin@uliege.be}
\affiliation{Space sciences, Technologies and Astrophysics Research (STAR) Institute, Universit\'{e} de Li\`{e}ge, B\^{a}t B5A, Sart Tilman, 4000 Li\`{e}ge, Belgium}

\begin{abstract}

Multi-Higgs models equipped with global symmetries produce scalar dark matter (DM) candidates stabilized 
by the unbroken symmetry. It is remarkable that a conserved $CP$ symmetry can also stabilize
DM candidates, provided it is a $CP$ symmetry of order higher than two.
CP4 3HDM, the three-Higgs-doublet model with $CP$ symmetry of order 4, is the simplest example of this kind.
It contains two mass-degenerate scalar DM candidates $\varphi$ and $\bar\varphi$, 
each of them being a CP4 eigenstate and, therefore, its own antiparticle. 
A novel phenomenological feature of this model is the presence 
of $\varphi\varphi \leftrightarrow \bar\varphi\bar\varphi$ conversion process, which conserves $CP$.
It offers a rare example of DM models in which self-interaction in the dark sector can significantly affect cosmological 
and astrophysical observables.
Here, we explore the thermal evolution of these DM species in the asymmetric regime.
We assume that a mechanism external to CP4 3HDM produces an initial imbalance of the densities of 
$\varphi$ and $\bar\varphi$. As the Universe cools down,
we track the evolution of the asymmetry through different stages,
and determine how the final asymmetry depends on the interplay 
between the conversion and annihilation $\varphi\bar\varphi \to $ SM 
and on the initial conditions. We begin with the analytic treatment of Boltzmann equations, 
present a detailed qualitative description of the process, and then corroborate it with numerical results
obtained using a dedicated computer code.
Finally, we check if the model can produce an observable indirect detection signal.

\end{abstract}

\maketitle

\section{Introduction}

\subsection{Dark matter candidates from the Higgs sector}

Many astrophysical and cosmological observations offer a compelling evidence of the existence 
of Dark Matter (DM) \cite{Hinshaw:2012aka,Ade:2015xua}.
From the particle physics perspective on DM, the main paradigm is to assume that the DM candidate is stable on
cosmological time scales, cold, i.e. non-relativistic at the onset of
galaxy formation, non-baryonic, neutral, and weakly interacting, and
that its relic abundance is obtained via thermal freeze-out in the early Universe, 
see e.g. \cite{Bertone:2004pz} for a review. 
Such particles are absent in the Standard Model (SM).
Together with the absence of an equally compelling evidence for scattering of DM on nuclei
\cite{Akerib:2016vxi,Aprile:2017iyp} or for astrophysical signals which would unequivocally confirm 
DM annihilation processes \cite{TheFermi-LAT:2017vmf},
it motivates the vast field of building models beyond the SM which include DM candidates.

One popular class of such models is represented by symmetry-based multi-Higgs models,
in which DM candidates arise naturally \cite{Ivanov:2017dad,Ilnicka:2018def}. 
If a model features inert scalars --- that is, fields not involved in generation of masses of 
gauge bosons or fermions --- and if the inert scalar sector transforms non-trivially
under a new global symmetry, which remains unbroken upon
minimization of the scalar potential, then the lightest inert scalar 
is automatically protected against decay and serves as a DM candidate. 
One of the simplest and actively studied examples is the Inert Doublet Model (IDM),
the version of the two-Higgs Doublet Model (2HDM) with the exact $\Z_2$ global symmetry
\cite{Deshpande:1977rw,Cao:2007rm,Barbieri:2006dq,LopezHonorez:2006gr}
which leads to interesting astrophysical and collider signatures
\cite{Ilnicka:2015jba,Belyaev:2016lok,Belyaev:2018ext,Kalinowski:2018ylg,Kalinowski:2018kdn}. 

Scalar DM models with more than two Higgs doublets have also attracted much attention.
In particular, within three-Higgs-doublets models (3HDMs), 
one can incorporate several global symmetry groups \cite{Ivanov:2012ry,Ivanov:2012fp},
which lead to a variety of options in DM model building \cite{Keus:2014jha,Ahriche:2015mea,Keus:2015xya,Cordero-Cid:2016krd}.
It is interesting to remark that in almost all multi-Higgs models it is extremely difficult
to probe and constrain self-interaction within the dark sector.
It can be rather strong and can bear certain patterns, but usually it does not affect the DM relic density 
nor can it be probed through the direct or indirect detection experiments, see for example a detailed analysis 
in the case of IDM \cite{Sokolowska:2011sb,Sokolowska:2011aa}.

\subsection{Dark matter stabilized by an exotic $CP$ symmetry}

In \cite{Ivanov:2015mwl}, another 3HDM was described, in which the only symmetry imposed 
was the $CP$ symmetry of order 4 (CP4). 
It is the first model featuring such a novel, physically distinct \cite{Haber:2018iwr} form of $CP$ symmetry 
without producing any accidental symmetries.
If CP4 remains unbroken after minimization of the Higgs potential,
the model leads to a pair of mass-degenerate scalar DM candidates,
which are stabilized, remarkably, by the $CP$ invariance in the scalar sector. 
This non-standard stabilization mechanism may have peculiar consequences
for the DM evolution, and this is what we explore in the present work.

Let us briefly describe the setting of this model, leaving subtle details for the \hyperref[appendix-CP4-3HDM]{Appendix}.
CP4 3HDM is a three-Higgs-doublet model based on Higgs doublets $\phi_{1,2,3}$ equipped 
with a generalized $CP$ symmetry of order 4 denoted CP4 \cite{Ivanov:2011ae,Ivanov:2015mwl}.
There are no new fields apart from the new Higgs doublets.
Its potential can be written as 
\bea
V &=& - m_{11}^2 \phi_1^\dagger \phi_1 - m_{22}^2 \left(\phi_2^\dagger \phi_2 + \phi_3^\dagger \phi_3\right)
+ \lambda_1 (\phi_1^\dagger \phi_1)^2 + \lambda_2 \left[(\phi_2^\dagger \phi_2)^2 + (\phi_3^\dagger \phi_3)^2\right]
\nonumber\\
&+& \lambda_3 \phi_1^\dagger \phi_1 \left(\phi_2^\dagger \phi_2 + \phi_3^\dagger \phi_3\right)
+ \lambda'_3 (\phi_2^\dagger \phi_2) (\phi_3^\dagger \phi_3) + 
\lambda_4 \left(|\phi_1^\dagger \phi_2|^2 + |\phi_1^\dagger \phi_3|^2\right)
+ \lambda'_4 |\phi_2^\dagger \phi_3|^2 \nonumber\\
&+& {\lambda_6 \over 2} \left[(\phi_2^\dagger\phi_1)^2 - (\phi_1^\dagger\phi_3)^2\right] +
\lambda_8(\phi_2^\dagger \phi_3)^2 + \lambda_9(\phi_2^\dagger\phi_3)\left(\phi_2^\dagger\phi_2-\phi_3^\dagger\phi_3\right) + h.c.
\label{V}
\eea
Here, all parameters apart from $\lambda_{8,9}$ are real.
This potential is invariant 
under the CP4 transformation $\phi_i(\vec r, t) \mapsto X_{ij} \phi_j^*(-\vec r, t)$, where matrix $X$ is
\be
X =  \left(\begin{array}{ccc}
	1 & 0 & 0 \\
	0 & 0 & i \\
	0 & -i & 0 
\end{array}\right)\,. 
\label{CP4}
\ee

The potential $V$ is the most general potential compatible with CP4, and it represents the minimal model
which realizes this novel form of $CP$ symmetry.

Although multi-Higgs-doublet models offer a large basis change freedom,
there exists no basis choice in which CP4 would take the ``standard'' form $\phi_i \mapsto \phi_i^*$.
Or equivalently, there is no basis in which all parameters of the potential would become real.
Technically, this is so because applying CP4 twice results not in the identity transformation but in a family symmetry 
$X X^* = \mathrm{diag}(1,\,-1,\,-1)$. One needs to apply CP4 four times to get the identity transformation.
Thus, CP4 is a genuinely novel form of $CP$ symmetry, physically distinct from the ``standard'' $CP$.
This distinction was also established at the level of observables \cite{Haber:2018iwr}.

In a detailed scan of the parameter space of this model, one would need to satisfy the boundedness-from-below conditions
and, to make the perturbative calculation reliable, the perturbative unitarity constraints. 
The necessary and sufficient boundedness-from-below conditions are not yet known,
but a simple set of sufficient conditions was proposed in Ref.~\cite{Ferreira:2017tvy}.
The perturbative unitarity constraints for CP4 3HDM were written and studied in \cite{Bento:2017eti,Kopke:2018vyw}.
However, since we do not perform a detailed scan in this work, we will not use these conditions.

We are interested in the regime where CP4 is preserved in the vacuum and protects the scalar DM candidates against decay.
The only vacuum expectation value (vev) alignment which conserves CP4 is $\lr{\phi_1^0,\phi_2^0,\phi_3^0} = (\varv,0,0)/\sqrt{2}$.
The vev symbol $\varv=246$ GeV is not to be confused with the velocity $v$.
We expand the doublets around this vacuum as
\be
\phi_1 = \doublet{G^+}{{1 \over \sqrt{2}}(\varv + h_{125} + i G^0)}\,,\quad
\phi_2 = \doublet{H^+_2}{{1 \over \sqrt{2}}(H + i a)}\,, \quad
\phi_3 = \doublet{H^+_3}{{1 \over \sqrt{2}}(h + i A)}\,, \label{fields-basis}
\ee
where $G^+$ and $G^0$ represent the would-be Goldstone bosons
which are absorbed into the gauge bosons $W^+$ and $Z$ by the Brought-Englert-Higgs mechanism.

The SM-like Higgs boson comes from the first doublet and has mass
$m_{h_{125}}^2 = 2\lambda_1 \varv^2 = 2m_{11}^2$. All other scalars coming
from the two inert doublets show a pairwise degenerate spectrum: 
\bea
m^2 \equiv m_{h,a}^2 &=& {1\over 2}\varv^2 \lambda_{346} - m_{22}^2\,,\nonumber\\
M^2 \equiv m_{H,A}^2 &=& {1\over 2}\varv^2 \bar{\lambda}_{346} - m_{22}^2
= m^2 + \lambda_6 \varv^2\,,\nonumber\\ 
m_{H^\pm}^2 \equiv m_{H^\pm_2, H^\pm_3}^2 
&=& {1\over 2}\varv^2 \lambda_3 - m_{22}^2 = m^2 + {1\over 2}\varv^2 (\lambda_6 -\lambda_4)\,,\label{inert-masses}
\eea
where we introduced the shorthand notation $\lambda_{346} = \lambda_3 + \lambda_4 - \lambda_6$
and $\bar{\lambda}_{346} = \lambda_3 + \lambda_4 + \lambda_6$.
Notice that the pair-degenerate mass spectrum is imposed by a discrete, not continuous symmetry \cite{Haber:2018iwr}.
Due to this degeneracy, the spectrum of the model resembles a typical 2HDM spectrum:
three values of neutral Higgs masses and one value of charged Higgs mass.
However, the counting of additional scalars is different from 2HDM, which can be detected in principle
by different production and decay rates, see more discussion in \cite{Haber:2018iwr}.

The two neutral fields $h$ and $a$ are the DM candidates protected by CP4 symmetry.
The relation $m < M$ implies $\lambda_6 > 0$ so that $\lambda_{346} < \bar{\lambda}_{346}$.
Also, we assume that $\lambda_6 - \lambda_4 > 0$ to guarantee that the lightest inert scalars are neutral.
The heavier inert scalars become unstable decaying into the lightest ones and into the SM fields.
They do not directly affect the relic density evolution.

Although $CP$ is conserved, one cannot classify the neutral scalars $h$ and $a$ as $CP$-even or $CP$-odd.
They are not even $CP$-eigenstates, as they transform under CP4 as $h \toCP -a$ and $a\toCP h$. 
However, one can combine them into CP4 eigenstates \cite{Ivanov:2015mwl,Aranda:2016qmp}:
\be
\varphi = {1\over \sqrt{2}}(h + i a)\,, \quad \varphi \toCP i\varphi\,.\label{J-eigenstates}
\ee
The fields $\varphi$ and $\varphi^*$ are $CP$-eigenstates 
but they are characterized by a $CP$-charge defined modulo 4 rather than $CP$-parity.
In order for this construction to be valid, the fields $h$ and $a$ cannot possess any 
other conserved quantum numbers, see \cite{Aranda:2016qmp} for a detailed discussion.

\subsection{CP4 3HDM in the asymmetric DM regime}

A first discussion of the qualitative features of the DM sector in CP4 3HDM was given in \cite{Ivanov:2015mwl,Haber:2018iwr}.
A more detailed quantitative insight into the DM evolution was given in \cite{Kopke:2018vyw}.
In this paper, we present a detailed exploration of another aspect of DM
in CP4 3HDM --- its evolution in the asymmetric DM regime.

The thermal evolution of asymmetric DM has been the subject of numerous studies, 
see \cite{Graesser:2011wi,Iminniyaz:2011yp,Zurek:2013wia,Petraki:2013wwa} and references therein.
The main idea is that, at some point in the thermal history of the Universe,
an asymmetry between DM particles and antiparticles was created, 
probably along with the baryon asymmetry of the Universe.
Then one has two main options.
If the DM particles $\varphi$ are associated with a new conserved global or gauge $U(1)$ charge,
then the asymmetry $n_\varphi - n_{\bar \varphi}$ remains unchanged in the later evolution,
which typically leads to an almost complete domination of DM particles over antiparticles.
In terms of the relative asymmetry $\delta$ defined as
\be
\delta = \fr{n_\varphi - n_{\bar \varphi}}{n_\varphi + n_{\bar \varphi}}\,,
\ee
this regime corresponds to $\delta \to 1$ at present.
A typical feature of these models is a very suppressed or absent present-day annihilation signal.
If the DM antiparticles have already died out, the DM particles cannot annihilate anymore.
Another possibility is that there exists a novel interaction which violates the DM number conservation.
If such violation is present in the quadratic terms, it will induce oscillations between DM particles and antiparticles
and will tend to wash out surviving asymmetry, unless the oscillations are extremely weak 
\cite{Buckley:2011ye,Tulin:2012re,Cirelli:2011ac}.

We show in this paper that the asymmetric DM evolution in CP4 3HDM has characteristic features
which lead to a richer list of options. 
The lightest inert particles in CP4 3HDM are $\varphi \equiv \varphi^*|0\rangle$ and $\bar\varphi \equiv \varphi|0\rangle$, 
which have equal masses but which are {\em not} particle and antiparticle of each other.
Each of them is an eigenstate of the $CP$ symmetry, since the conserved symmetry CP4 acts on the complex fields 
$\varphi$ and $\varphi^*$ as
\be
\varphi \toCP i \varphi,\quad \varphi^* \toCP -i \varphi^*\,,\label{CP-property}
\ee
see more details in the \hyperref[appendix-CP4-3HDM]{Appendix}. Thus, the asymmetric DM regime in CP4 3HDM
is {\em not} directly related to the particle-antiparticle asymmetry.
It is rather an asymmetry between two sorts of mass-degenerate complex fields. 

Having said that, we admit that it is easier to discuss the properties of $\varphi$ and $\bar\varphi$ 
by calling them particle and antiparticles, respectively. This is a language abuse, but we will stick to it in this paper.
We stress once again that, despite using this language, we recognize that $\varphi$ and $\bar\varphi$ are their own antiparticles.

The transformation law under CP4 given by Eq.~\eqref{CP-property} allows us to label
the two particles not with $CP$ parities but with $CP$ charges $q$ defined modulo 4.
This $CP$ charge protects them against decay and against $\varphi \leftrightarrow \bar\varphi$, 
as well as against annihilation $\varphi\varphi \to Z$.\footnote{If the heavier inert states are taken into account,
there appears a number of coannihilation channels. However, if these scalars are significantly heavier, they
quickly decay and do not affect the thermal freeze-out process nor the late-time evolution.} 
However, the scalar potential contains terms that induce the {\em conversion} process 
(see the \hyperref[appendix-CP4-3HDM]{Appendix} for the conversion cross section):
\be
\varphi \varphi \longleftrightarrow \bar\varphi \bar\varphi \,,
\ee
which is well compatible with $CP$ charge conservation: $+2$ is equal to $-2$ modulo 4.
Therefore, although the DM oscillations are forbidden, $\bar\varphi$'s can still be regenerated in $\varphi$-dominated Universe,
pick up a $\varphi$ and annihilate into SM fields.
Notice that this conversion is governed exclusively by the quartic self-interaction in the DM sector.

The relative intensity of the conversion and annihilation processes will affect the present day relic asymmetry as well as
the total DM abundance. Depending on the parameters of the model and on the initial asymmetry\footnote{%
We do not discuss the origin of this primordial asymmetry and its possible relation with the baryon asymmetry of the Universe. 
It certainly goes beyond CP4 3HDM {\em per se} and requires additional phenomena, 
which may exist in a more comprehensive model at higher energy and leads to CP4 3HDM at the electroweak scale.
In this work, we simply postulate an initial asymmetry at certain temperature and then investigate its evolution
during thermal freeze-out.}, the outcome of the thermal evolution may be almost fully symmetric, almost total asymmetric,
or partially asymmetric. We will trace the boundary between the extreme regimes, link the final asymmetry with its primordial value,
and investigate the prospects of the indirect detection signal in these regimes.
Thus, this model offers a rare example where self-interaction in the DM sector does significantly affect the cosmological observables.
 
The structure of the paper is as follows. In Section~\ref{section-restoration} we write down Boltzmann equations
and discuss the qualitative features of their solutions. 
We show that, starting from the relatively low values of $x \equiv m/T \sim 1$, the relative asymmetry $\delta$
first drops (depletion stage), then steadily grows (recovery stage), and either approaches unity or flattens at some value.
In Section~\ref{num_an}, we corroborate these results with the numerical analysis performed by means of a dedicated evolution code.
Then, in Section~\ref{indirect}, we estimate the annihilation signal in various regimes,
and finally we draw our conclusions. 
The \hyperref[appendix-CP4-3HDM]{Appendix} provides some details of CP4 3HDM and the properties of its DM candidates.

\section{Asymmetry evolution: qualitative analysis}\label{section-restoration}

\subsection{Boltzmann equations}

The evolution of particles $\varphi$ and $\bar{\varphi}$  (which, in abuse of language, will often be called below particle and antiparticle)
in the homogeneous and isotropic expanding Universe is described by a system of coupled Boltzmann equations
\bea
\begin{cases}
	\derivX{Y} = - A(x) \left[Y \bar{Y} - \Yeq^2(x)\right] - B(x) \left(Y^2 - \bar{Y}^2\right)/4\\[1mm]
	\derivX{\bar{Y}} = - A(x) \left[Y \bar{Y} - \Yeq^2(x) \right] + B(x) \left( Y^2 - \bar{Y}^2 \right)/4 \, ,
\end{cases}
\label{Boltzmann_initial}
\eea
where the factor $1/4$ is introduced for future convenience.
In these equations, $Y$ and $\bar{Y}$ denote the number density in the comoving volume of $\varphi$ and $\bar{\varphi}$, respectively. 
For the sake of brevity, from now on we refer to these quantities as ``densities'', except for Section \ref{indirect}, where we will need to deal with physical number densities. The dot above a symbol denotes the derivative with respect to $x$, e.g. $\dot{Y} \equiv dY/dx$, where $x = m/T$, 
$m$ is the mass of $\varphi$ and $T$ is the temperature of the plasma. Eqs.~\eqref{Boltzmann_initial} explicitly mention the $x$ dependence 
of the parameters; of course, $Y$ and $\bar Y$, being variables, also depend on $x$, which is implicitly assumed in all equations.
$\Yeq(x)$ is the equilibrium density at given $x$ of both
$\varphi$ and $\bar\varphi$ in the $\varphi/\bar\varphi$ symmetric plasma. In an asymmetric situation, the actual equilibrium densities
of $\varphi$ and $\bar\varphi$ differ but their product is still equal to $\Yeq^2$.
Finally, $A(x)$ and $B(x)/4$ denote the intensities of the annihilation $\varphi\bphi \leftrightarrow \SM$ and conversion $\varphi\varphi \leftrightarrow \bphi\bphi$ processes, respectively. The first one is determined as follows\footnote{The details on the derivation of the exact velocity-averaged cross sections and equilibrium number density can be found, for instance, in the Appendix of \cite{Ahmed:2017dbb}.}
\be
A(x) = \sqrt{\frac{g(x)\pi}{45}}\frac{mM_{\rm Pl}}{x^2} \svSM(x) \, ,
\label{Intensity}
\ee
where $g$ denotes the number of relativistic degrees of freedom at the moment $x$, $M_{\rm Pl}$ is the Planck mass, 
and $\svSM$ is the total velocity-averaged cross section of all interactions that turn $\varphi$ and $\bphi$ into SM particles. 
A similar expression holds for $B(x)$, where one uses the corresponding velocity-averaged cross section $\svC$ instead.
Notice that $A(x)$ and $B(x)$ are expected to show distinct $x$ dependences.
In the case of $s$-wave annihilation of slow DM particles into lighter SM fields,
$\svSM(x) \approx$ const. 
For the $\varphi\varphi \leftrightarrow \bphi\bphi$ conversion process, due to $m_\varphi = m_{\bar\varphi}$, 
it is the cross section $\sC$ that is approximately constant, see Eq.~\eqref{sigma-conv}. 
Thus, taking into account the relation between the average velocity and temperature \cite{Gondolo:1990dk,Cannoni:2013bza}, the velocity-averaged cross section is $\svC = \sC \lr{v} = \sC \cdot 2/\sqrt{\pi x}$. Finally, we arrive at the following expression
\be
B(x) = \sqrt{\frac{32 g(x)}{45}}\frac{mM_{\rm Pl}}{x^{5/2}} \sC \, .
\label{conversion-intensity}
\ee
Neglecting the slow change of $g$ with $x$, one can get the following approximation
\be
A(x) \approx \frac{a}{x^2} \, , \qquad B(x) \approx \frac{b}{x^{5/2}} \, ,
\ee
which shows that the relative importance of the conversion process to annihilation to SM fields 
decreases as $x$ grows. 

Let us now split the two densities into the symmetric and asymmetric contributions:
\be
\Yt = \frac{Y+\bar{Y}}{2} \, , \qquad \dY = \frac{Y-\bar{Y}}{2} \, , \qquad \delta = \frac{\dY}{\Yt} \, .
\ee
Here, $\Yt$ stands for half the total density of $\varphi$ and $\bphi$, 
which we will call for brevity just the total density, $\dY$ stands for half their difference, 
while $\delta$ quantifies the relative amount of asymmetry between $\varphi$ and $\bphi$. 
The fully symmetric case corresponds to $\delta = 0$, while the complete asymmetry implies $\delta = 1$ or $-1$. 
For concreteness, we will assume that $\varphi$ dominate over $\bar\varphi$ so that the asymmetry is always non-negative.

By taking the sum and the difference in \eqref{Boltzmann_initial}, one can recast these equations into
evolution of $\Yt$ and $\dY$: 
\bea
\begin{cases}
	\derivX{\Yt} = - A(x) \left[\Yt^2 - \dY^2 - \Yeq^2(x)\right]\\[1mm]
	\derivX{\dY} = - B(x)\Yt\dY\,.
\end{cases}
\label{Boltzmann_asym-1}
\eea
One sees that the conversion process washes out the {\em absolute value} of the asymmetry.
However, since the total density also drops, the evolution of the {\em relative} asymmetry may be quite sophisticated.
To describe it, we further recast \eqref{Boltzmann_asym-1} into equations for $Y_t$ and $\delta$:
\bea
\begin{cases}
	\derivX{\Yt} = - A(x) \left[\Yt^2 (1 - \delta^2) - \Yeq^2(x)\right]\\[1mm]
	\derivX{\delta} = - \delta \left[B(x)\Yt + \dot{\Yt}/\Yt\right]\,.
\end{cases}
\label{Boltzmann_asym-2}
\eea
The last equation can also be explicitly expanded as
\be
\derivX{\delta} = \delta \cdot Y_t \left[A(x)\left(1 - \delta^2 - \frac{\Yeq^2}{\Yt^2}\right) - B(x)\right]\,.
\label{Boltzmann_asym-3}
\ee
One observes that there are several competing terms inside the brackets of \eqref{Boltzmann_asym-3}.
Thus, the relative asymmetry can grow, decrease, or change the regime during the $x$ evolution. 
It is this evolution that we will investigate below.

\subsection{Qualitative analysis without conversion}

Before presenting numerical results, let us discuss the expected qualitative features of the $x$ evolution of the relative asymmetry.
We begin with the case when no conversion takes place, $B(x) = 0$. 
In this case the evolution of the relative asymmetry $\delta$ can still be non-trivial, 
since the bracket in Eq.~\eqref{Boltzmann_asym-3} can, in principle, change sign.
Although the main features of the evolution are well known in this case, see e.g. \cite{Graesser:2011wi},
we outline them to set up the notation.

Suppose one starts at $x = x_0 = {\cal O}(1)$ with a non-zero initial $\delta_0$ and some initial density $Y_{t0}$. 
These initial conditions, $\delta_0$ and $Y_{t0}$, must be considered as free parameters,
since the origin and the amount of the primordial asymmetry go beyond CP4 3HDM and 
need to be derived from a theory at a higher mass scale, which simplifies to CP4 3HDM at the electroweak scale.

At first, when $x$ is well below the freeze-out point for $Y_t$, 
the total density $\Yt$ quickly approaches the equilibrium density $\Yeq/\sqrt{1-\delta^2}$, either from above or from below, 
depending on $\delta_0$ and $Y_{t0}$.
In particular, if we start with $\delta_0$ close to 1 and $Y_{t0} \sim \Yeq$,
we can see a brief drop in $\delta$ due to a momentary increase of $Y_t$.
This equilibration episode, which is driven by the choice of the initial conditions, will be quickly taken over by the steady growth of $\delta$.

The subsequent evolution of $\delta$ can be deduced directly from \eqref{Boltzmann_asym-2}. 
If $\delta_0 \ll 1$ and $\delta$ stays small during thermal evolution, 
then, well before the freeze out, $\Yt \approx \Yeq$ and is approximately determined by the Maxwell-Boltzmann distribution: 

\be
\Yeq(x) = c(x) \, x^{3/2}\exp(-x) \, , \quad c(x) = {90 \over g(x) (2\pi)^{7/2}}\,.
\label{equil_dens}
\ee
Plugging this into the second equation in \eqref{Boltzmann_asym-2} one immediately gets
\be
\derivX{\delta} \approx \delta \left(1 - \frac{3}{2x}\right) \, .
\label{asym_change_sign}
\ee
For values of $x \gtrsim 10$, it further simplifies to $\derivX{\delta} \approx \delta$ and results in a very simple dependence:
\be
\delta(x) \propto \exp(x)\,.
\label{delta_restore}
\ee
If $\delta$ still stays small during this entire evolution until the freeze-out point for the total density $x= x_f$, 
one obtains the growth factor of the order of $\exp(x_f-x_0)$.

After the freeze-out for the total density, $Y_t$ reaches a plateau, and $\Yeq/\Yt$ quickly drops and can be neglected. 
If $\delta$ is still considerably less than $1$, the first equation in \eqref{Boltzmann_asym-2} can be explicitly integrated 
and one gets
\be
Y_t(x\to\infty) = \Yt(x_f)\left(1 + a \fr{\Yt(x_f)}{x_f}\right)^{-1} \approx {x_f \over a}\,.
\label{Yt-flattening}
\ee
In the last passage here, we took into account that the extra term in the brackets is typically ${\cal O}(10) \gg 1$.
Just like in the usual symmetric case, the freeze-out moment $x_f \sim 25$ 
is set by the observed relic density.
Since $\Delta = \mathrm{const}$ in the absence of conversion, we have $\delta \propto \Yt^{-1}$,
and the late-time value of the asymmetry $\delta_{\infty}$ grows from $\delta(x_f)$
by the same factor as in Eq.~\eqref{Yt-flattening}. 
Thus, it is not guaranteed that $\delta$ eventually reaches unity:
if the initial value $\delta_0$ was sufficiently small, $\delta$ can stay small throughout the entire freeze-out process.

It may also happen that $\delta$, being initially moderately small, grows close to unity during the above evolution. 
In this case, the $1-\delta^2$ factor in \eqref{Boltzmann_asym-2} changes the dynamics of $\Yt$
and shifts the freeze-out moment $x_f$ to earlier times. 
One can estimate the moment when $\delta$ approaches unity and the total density effectively freezes out. 
For that, we restore the non-zero $\delta$ and, during the thermal equilibrium stage, 
write the total density as $\Yeq/\sqrt{1-\delta^2}$. The ratio $\dot{\Yt}/\Yt$ will now contain contributions
not only from $\Yeq$ but also from $\delta$. Substituting it again in the second equation of \eqref{Boltzmann_asym-2} and omitting the $3/2x$ term for simplicity, we obtain after some algebra:
\be
\derivX{\delta} \approx \delta \left(1 - \delta^2 \right) \, .
\label{asym_change_sign_mod}
\ee
Integrating the above equation, we get 
\be
\delta(x) \approx \frac{\delta_0 \exp(x-x_0)}{\sqrt{1 + \delta_0^2 \exp(2x-2x_0)}} \, ,
\ee
which shows how $\delta(x)$ approaches unity.

For a rough estimation, let us {\em define} the freeze-out moment for the total density
as the moment at which $\delta = 0.99$. We then obtain
\be
x_f \approx 1 + \ln \left(\frac{7}{\delta_0}\right) \, ,
\ee
where for definiteness we set the initial moment $x_0=1$.
We see that any sizable initial asymmetry $\delta_0 > 10^{-8}$ shifts 
the total density freeze out to $x_f \lesssim 22$, 
which is earlier than in the symmetric case.

Notice that the moment $x=x_f$ only settles $Y_t \approx \Delta$ in the case of the early freeze-out, 
while the density of antiparticles and, therefore, $\delta$ keep evolving.
In this regime, the relevant variable is not $\delta$ itself, but the relative amount of antiparticles:
\be
r \equiv {\bar Y \over Y + \bar Y} = {1 - \delta \over 2}\,.
\ee
For $x > x_f$, it still keeps decreasing exponentially, 
$r \approx 5\cdot 10^{-3} \exp[-2(x-x_f)]$, and it settles to $r(\infty)$ 
only when the evolution departs from thermal equilibrium.
The evolution equation for $r \ll 1$ takes the following form
\be
\dot{r} = -2 \Delta  A(x) \left(r - {\Yeq^2 \over 4\Delta^2}\right)\,.
\ee
Solving it and analyzing the solution with Laplace's method keeping in mind
that $a \Delta \gg 1$, one finds that $r$ begins flattening out
around $x \sim x_r = \sqrt{a\Delta}$, and its asymptotic value is exponentially suppressed as
\be
r(\infty) \approx {\sqrt{\pi}\over 2} a^2 c^2 (a\Delta)^{-1/4}\exp(- 4\sqrt{a\Delta})\,,\label{final-r}
\ee
where $c$ is the prefactor in the equilibrium density, Eq.~\eqref{equil_dens}.

It is important to appreciate that the value of $\Yt(\infty) \approx \Delta$ and the parameter $a$
are not related anymore as it was in Eq.~\eqref{Yt-flattening}.
One can, for example, consider the annihilation cross section parameter $a$ much larger 
than what is typically needed for a given mass in the symmetric DM scenario for the same mass. 
The final total density will still be very close to $2\Delta$, but the relative contribution
of antiparticles will be strongly suppressed according to \eqref{final-r}.

\subsection{Switching on conversion}\label{chapter_conversion}

\begin{figure}[h!]
	\begin{tikzpicture} [
	catg/.style={draw,dashed, minimum height=2em},
	thick/.style=      {line width=0.8pt}
	]
	
	\draw [thick][arrows={->[scale=2]}] (0,0) -- (15,0) ;
	\draw [thick](0,-0.1) -- (0,0.1);
	\draw [thick](4,-0.1) -- (4,0.1);
	\draw [thick](10,-0.1) -- (10,0.1);
	
	\node [text width=1cm, align=center] at (0, -0.4)  {$x_0$};
	\node [text width=1cm, align=center] at (4, -0.4)  {$x_*$};
	\node [text width=1cm, align=center] at (10, -0.4)  {$x_f$};
	\node [text width=1cm, align=center] at (15, -0.4)  {$x$};
	
	\node [text width=3cm, align=center] at (1.5, 0.4)  {Depletion stage};
	\node [text width=3cm, align=center] at (4, 0.4)  {$\Delta$ stabilizes};
	\node [text width=3cm, align=center] at (7, 0.4)  {Recovery stage};
	\node [text width=3cm, align=center] at (12.5, 0.4)  {Flattening stage};

	\node [text width=3cm, align=center] at (2, -0.4)  {$\delta\!\!\searrow$};
	\node [text width=3cm, align=center] at (7, -0.4)  {$\delta\!\!\nearrow$};
	\node [text width=3cm, align=center] at (12.5, -0.4)  {$\delta\!\!\nearrow$};
	
	\end{tikzpicture}
	\caption{Schematic diagram of the thermal evolution of the asymmetry $\delta$. The moment $x_0$ denotes the starting value of $x$, at which $\delta = \delta_0$, $x_*$ corresponds to the minimum of $\delta$, and $x_f$ is the point at which the total density freezes out. The scales are approximate. See the text below for further details.}
	\label{diagram}
\end{figure}
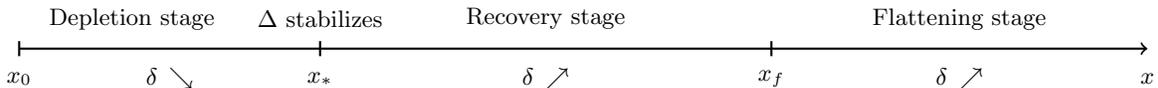
Let us now include the conversion process, $B(x) \not = 0$. Its role is to wash out asymmetry,
and the question is whether this washing out is sufficiently strong to modify the dynamics of the asymmetry. 
Since the conversion process is determined by a combination of quartic self-coupling constants in the dark sector (see \hyperref[appendix-CP4-3HDM]{Appendix})
this situation is the rare example of DM model where interactions in the dark sector have strong impact on the cosmological observables.
 
The first observation is that the conversion process may lead to a pronounced {\em depletion} of the asymmetry
at the early stage of evolution. The entire thermal history of the asymmetry may now have three distinct stages, 
schematically shown in Fig.~\ref{diagram}. At the beginning of the evolution, the initial value of 
$\delta$ drops down due to the conversion ({\em depletion stage}). Then, while the density and the conversion cross section decrease, the conversion process becomes ineffective and $\delta$ reaches its minimum at $x_*$ and starts to grow ({\em recovery stage}). This is because the equilibrium symmetric part of the total density decreases due to the expansion. After the total density freezes out the proportion of particles and antiparticles ceases to change as well and $\delta$ approaches its constant value ({\em flattening stage}). Note, that the beginning of the recovery stage also corresponds to the moment, when the value of the absolute asymmetry $\Delta$ becomes constant.

If, during the depletion stage, the two processes are of the same intensity, $B \sim A$, 
the $B(x)$ term in \eqref{Boltzmann_asym-3} strongly dominates and forces $\delta$ to sharply decrease
all the way to the freeze-out moment, skipping the recovery stage and leading to a vanishingly small final asymmetry.
If one wants the asymmetry to recover, one must assume that $B$ is not only smaller, but significantly smaller than $A$.
Indeed, during the depletion stage, the coefficient $A$ stands in front of $1-\delta^2 - \Yeq^2/\Yt^2 \ll 1$,
and the conversion process competes with this strongly reduced annihilation.
As $x$ grows, $\delta$ decreases and $\Yt$ starts to depart from $\Yeq$.
The coefficient $1-\delta^2 - \Yeq^2/\Yt^2$ steadily grows until the moment $x=x_*$ when 
the bracket in \eqref{Boltzmann_asym-3} changes sign, $\delta$ reaches its minimum value $\delta_*$, and the recovery stage begins.

As long as $\delta$ stays small $\Yt \approx \Yeq$, one can explicitly integrate the second equation in \eqref{Boltzmann_asym-2} to obtain the following behavior: 
\be
\delta(x) \approx \delta_0 \exp\left[(x-x_0) - m\tilde{M}_{\rm Pl}\sigma_{\rm conv} \Big({\rm E_1}(x_0) - {\rm E_1}(x)\Big) - \fr{3}{2} \ln(x/x_0)\right] \, .
\label{delta_depletion}
\ee
To keep the expressions compact, we introduce here $\tilde{M}_{\rm Pl} = \sqrt{45/g \pi^7}\,  M_{\rm Pl}$,
which differs from the commonly used $M^*_{\rm Pl}$ defined through the relation $H(T) = M^*_{\rm Pl}/T^2$ as 
$\tilde{M}_{\rm Pl} = 2 M^*_{\rm Pl}/\pi^2$. 
Assuming that $g \sim 100$ and does not change substantially during the thermal evolution, 
we estimate $\tilde{M}_{\rm Pl} \approx 10^{-2} M_{\rm Pl} \approx 10^{17}$~GeV.
Finally, ${\rm E_1}(x)$ is the exponential integral defined as
\be
E_1(x) = \int_x^\infty {dt \over t} e^{-t}\,,
\ee 
which can be approximated by $E_1(x) \approx e^{-x}/x$ already for moderately large values of $x$.

The depletion and recovery stages arise from the interplay of the first two terms
in the exponential of Eq.~\eqref{delta_depletion}.
At moderate values of $x$, the sharply decreasing second term
drives $\delta$ down. 
We highlight here the fact that the magnitude of the drop in $\delta$ 
is exponentially sensitive to the value of the conversion cross section.
When this drop flattens, the growing first term takes over and 
leads, in the recovery stage, to a steady growth of $\delta$ similar to Eq.~\eqref{delta_restore}.

The position of the minimum $x_*$ can be read off Eq.~\eqref{delta_depletion}.
Neglecting the slowly varying $\ln(x/x_0)$ term,
one obtains a transcendental equation for $x_*$:
\be
m \tilde{M}_{\rm Pl} \,\sigma_{\rm conv} \, e^{-x_*} = x_*\,.
\label{minimum_cond_transc}
\ee
To the logarithmic accuracy,  we can estimate the solution as
\be
x_* = \ln\left(\frac{m\tilde{M}_{\rm Pl} \,\sigma_{\rm conv}}{x_*}\right) 
\approx \ln\left(m\tilde{M}_{\rm Pl} \,\sigma_{\rm conv}\right) \, .
\label{minimum-condition}
\ee
We could get the same expression just by looking at the moment 
when the \textit{absolute value} of the asymmetry $\Delta$ freezes out, 
analogously to the common freeze out of the DM density in the symmetric case\footnote{See Section 9.2 in \cite{gorbunov2011introduction} for the derivation of the formula, similar to Eq.~\eqref{minimum-condition}.}. Indeed, during the recovery stage, the conversion process effectively switches off,
and it happens slightly after $x_*$. The value of $\Delta$ at this point can be estimated as
\be
\Delta_* \approx \Delta_0 \exp \left[-m\tilde{M}_{\rm Pl}\sigma_{\rm conv} \left({\rm E_1}(x_0) - {\rm E_1}(x_*)\right)\right]
\sim \Delta_0 \exp \left[-m\tilde{M}_{\rm Pl} \sigma_{\rm conv}  {\rm E_1}(x_0)\right]\, ,
\ee 
where we neglected ${\rm E_1}(x_*) \ll {\rm E_1}(x_0)$ for $x_* \gtrsim 3$.

The subsequent evolution depends on the value of $\delta$ at this moment.
If $\delta$ stays small, one expects approximately the same exponential grows of $\delta$ 
as in Eq.~\eqref{delta_restore}. 
At $x=x_f$, the moment of freeze-out for the total density, $\delta$ increases with respect to $\delta_*$
by the factor $\exp(x_f - x_*)$. If $\delta$ still remains small at this moment, the flattening stage unfolds, during which 
$\delta$ experiences an additional residual increase by the same factor as in Eq.~\eqref{Yt-flattening}.
Alternatively, if $\delta$ approaches unity, then the antiparticle contribution dies out similarly
to the previous case, with $\Delta$ being effectively frozen at $x\approx x_*$.

\subsection{Locating the boundary}

We showed above that there are two extreme regimes: weakly asymmetric, with the final $\delta \ll 1$,
and strongly asymmetric, when $\delta \approx 1$ and $r \ll 1$ at the end of thermal evolution. 
Let us now locate the boundary between the two regimes.

We see that the stronger is the conversion cross section $\sigma_{\rm conv}$, 
the more pronounced is the drop and the later asymmetry begins to restore. 
If this happens rather close to the total density freeze out, 
the asymmetry does not have enough time to recover to any considerable value. 
One can estimate the value of $\sigma_{\rm conv}$ which can lead, starting from a small initial $\delta_0$, 
to a considerable asymmetry around the total density freeze out.
One just needs to extend Eq.~\eqref{delta_depletion} all the way to $x_f$
and require that the final result be of order 1:
\be
\ln \delta_0 + (x_f-x_0) - x_* \exp(x_*) {\rm E_1}(x_0) - \fr{3}{2} \ln(x_f/x_0) = 0 \,.
\label{x_min_condition}
\ee
Notice that we neglected here $E_1(x_f)$ with respect to $E_1(x_0)$.
For $x_0 = 1$, $\delta_0 = 0.1$ and $x_f \sim 20 \div 30$ we numerically obtain $x_* \sim 3$. 
It confirms that the conversion interaction should freeze out relatively early compared to the total density freeze out, 
in order for a sufficient asymmetry to survive at the end. 

Next, using Eq.~\eqref{minimum_cond_transc}, we can recast this condition into an upper bound of the 
conversion cross section:
\be
\sigma_{\rm conv} \lesssim \frac{x_* \exp(x_*)}{m\tilde{M}_{\rm Pl}}\,.
\label{sigmaV_conv_limit}
\ee
For the value of $x_*$, obtained above, and for $m \sim 100 \GeV$, this relation gives $\sigma_{\rm conv} \lesssim 10^{-17} \GeV^{-2}$. Comparing this result to the typical velocity-averaged cross section of the interaction between WIMPs and SM particles needed to reproduce the correct relic abundance 
$\sigma_{\rm SM, 0} = 2.5 \cdot 10^{-9} \GeV^{-2}$, one sees that the conversion cross section 
has to be about eight orders of magnitude smaller.
Within CP4 3HDM, the cross section of the conversion process $\varphi\varphi \leftrightarrow \bar\varphi\bar\varphi$ 
is given by Eq. \eqref{sigma-conv}. Thus, the upper bound on the conversion cross section \eqref{sigmaV_conv_limit}
can be rewritten as an upper bound on the quartic coupling:
\be
|\lambda_{\rm conv}| \lesssim 10^{-5}\quad \mbox{for $x_0 = 1$}\,.
\label{critical-lambda-estimate}
\ee
Note, however, that this constraint becomes less severe if the initial asymmetry is generated
at larger values of $x_0 > 1$, see also discussion in Section~\ref{num_an}. 
We conclude that the exact bound on $\lambda_{\rm conv}$ is very sensitive to the moment $x_0$
and cannot be predicted purely within CP4 3HDM.

If the value of $\sigma_{\rm conv}$ is small compared to the upper bound \eqref{critical-lambda-estimate}
and $\delta_0$ is not vanishingly small, then $\delta$ approaches unity during the recovery stage,
which modifies the freeze-out moment $x_f$ shifting it to smaller values. 
The relic density in this case is determined by the value of the absolute asymmetry $\Delta_*$
at the moment of the effective decoupling of the conversion processes, which also sets the relic density:
\be
\Omega h^2 \propto \Delta_* \approx \Delta_0 \exp \left[-m\tilde{M}_{\rm Pl} \sigma_{\rm conv}  {\rm E_1}(x_0)\right]\,.
\label{Omegah_early_freezeout}
\ee
This result is interesting for two reasons. First, similarly to the asymmetric case without conversion, we see that the relic density does not depend on the value of $\sigma_{\rm SM}$, given that it is not extremely small so that the symmetric freeze out would occur later than the asymmetric one. Second, from the model-building perspective, the presence of the conversion brings more parametric freedom, since the value of the relic density is not fixed by the initial conditions.

In summary, depending on the strength of the conversion process, 
we expect to observe two distinct regimes.
For sufficiently weak conversion, with the bound \eqref{critical-lambda-estimate} satisfied,
one expects that the initial asymmetry, evolving through a depleted intermediate stage,
will recover by the freeze-out moment $x_f$ and can eventually grow to values close to unity.
The resulting relative contribution of antiparticles can be extremely small 
if the annihilation into SM is effective so that the parameter $a\Delta_*$ is large enough, see Eq.~\eqref{final-r}.
For a stronger conversion process, the recovery is only partial.
Even if the conversion process by itself is very weak at the moment of freeze-out $x_f$,
it was stronger at earlier times and it shifted the position of the minimum $x_*$ to larger values.
As a result, $\delta$ did not have enough time to grow to unity, and its final value can be rather small.

\section{Asymmetry evolution: numerical analysis} \label{num_an}

Though there are several available codes for numerical computation of the relic abundance, including micrOMEGAs \cite{Belanger:2014vza}, DarkSUSY \cite{Bringmann:2018lay} and MadDM \cite{Ambrogi:2018jqj}, none of them is capable of handling asymmetric DM models with particle-antiparticle conversion processes. For that reason we have developed our own C++ code. 

\begin{figure}[t!]
	\center{\includegraphics[width=0.58\textwidth]{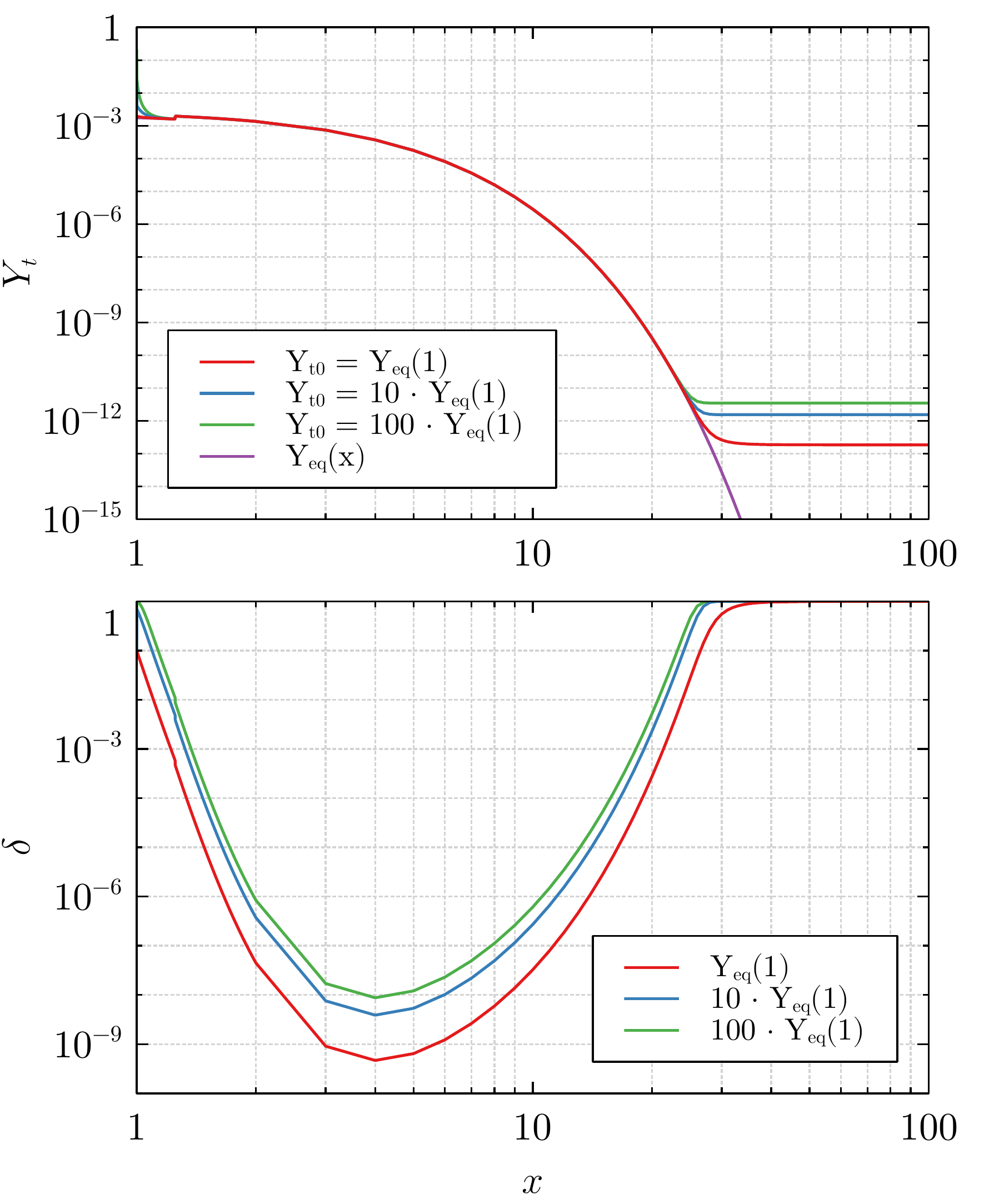}}
	\caption{Total density of DM $\Yt$ (\textit{top}) and asymmetry of DM $\delta$ (\textit{bottom}) as a function of $x$ for different values of the initial total DM density $Y_{t0}$. 
	The magenta curve in the \textit{top} plot shows the evolution of the equilibrium density $\Yeq(x)$. 
	The other initial conditions here 	have their default values; in particular, $\delta_0 = 0.1$. 
	The value of the conversion constant is $\lambda_{\rm conv} = 10^{-5}$. See the text below for more details.}
	\label{plot_Y_in} 
\end{figure}

We use two different integration methods at different stages of the thermal evolution: the implicit Euler method for $x \sim 1 \div 20$ and the classical Runge-Kutta method (RK4) for $x \gsim 20$. While the latter method has a better precision ${\cal O}(t^4)$ for a given integration step $t$, it fails to provide adequate solutions when the total density $\Yt$ is extremely close to the equilibrium density $\Yeq$. The backward Euler method is the simplest implicit method, which manages to solve the system \eqref{Boltzmann_asym-1} in this stiff regime. The global truncation error for this method is ${\cal O}(t)$, so one has to use very small integration steps to get sufficient precision. In order to validate the solution obtained with the backward Euler method, we applied the implicit midpoint method, which has an accuracy ${\cal O}(t^2)$. For $t \lesssim 10^{-3}$, the results of these two methods are consistent up to ${\cal O}(t)$.

We also compared our results in the symmetric case and in the absence of the conversion interaction to a simple freeze-out analytic approximation of the relic density \cite{gorbunov2011introduction}
\be
\Omega h^2 \approx \fr{3}{2}\cdot 10^{-10} g^{-1/2}\left(\fr{\GeV^2}{\sigma_0}\right) \ln \left(\fr{m M^*_{\rm Pl}\sigma_0}{(2\pi)^{3/2}}\right)\, ,
\ee
and to the results of micrOMEGAs. As an example, for $m = 250 \GeV$ and $\sigma_0 = 9.8 \cdot 10^{-8} \GeV^{-2}$ the formula above gives $\Omega h^2 = 0.22$. In the case of the Singlet DM model, where the values of the parameters correspond to $m$ and $\sigma_0$ given above, micrOMEGAs returns $\Omega h^2 = 0.16$, while our code gives $\Omega h^2 = 0.2$.

For our analysis of the asymmetry evolution we adopt the following default values of the model parameters: $m = 250 \GeV$, $\lambda_{346} = 0.2$, $\lambda_{3} = 0.9$, $\lambda_{\rm conv} = 10^{-5}$. In the model we consider, which is presented in detail in Appendix \ref{appendix-CP4-3HDM}, the second and the third parameters determine the intensity of the interaction with the SM sector. The parameter values presented above correspond to $\svSM \approx \sigma_{\rm SM} \sim 10^{-7} \GeV^{-2}$. Given that $\svSM$ is much larger than $\svC$, the last parameter in the list affects the asymmetry evolution the most, so our main goal is to study its variations.
 
When implementing the evolution of asymmetric DM, we need to fix the initial conditions:
that is, we need to select the starting point $x_0$ and select the initial total density $Y_{t0}$ and the initial asymmetry $\delta_0$.
Our default choice is $x_0 = 1$, $\delta_0 = 0.1$ and $Y_{t0} = \Yeq$.
This initial choice brings us from the start sufficiently close to equilibrium for the total density evolution.
As a stability check, we also performed runs in which $\delta_0 = 0.1$ 
but the initial total density was chosen to be $10\Yeq$ and $100\Yeq$,
which are very out-of-equilibrium situations. In all the cases, we observe (see Fig. \ref{plot_Y_in}) that the total density approaches the equilibrium density very fast, within $\Delta x \ll 1$. This quick equilibration of $Y_t$ also produces the very sharp jump of 
$\delta$ to the vicinity of one; although the three curves at the lower plot of Fig.~\ref{plot_Y_in} looks as if starting from different values of $\delta$, the initial value in all of them is $\delta_0 = 0.1$.
The small discontinuity of the curves visible in Fig.~\ref{plot_Y_in} at $x = 1.25$ is the artifact of our code, in which the number of relativistic degrees of freedom $g$ changes discretely with $x$ and one of these jumps takes place at $T = 200 \GeV$. This is the only plot, where the minimal step between points in the interval $x = \left[1,2\right]$ is taken to be $10^{-3}$ to show in detail the transition from the initial state to the equilibrium. 

\begin{figure}[t!]
	\begin{minipage}[h]{0.45\linewidth}
		\center{\includegraphics[width=1\textwidth]{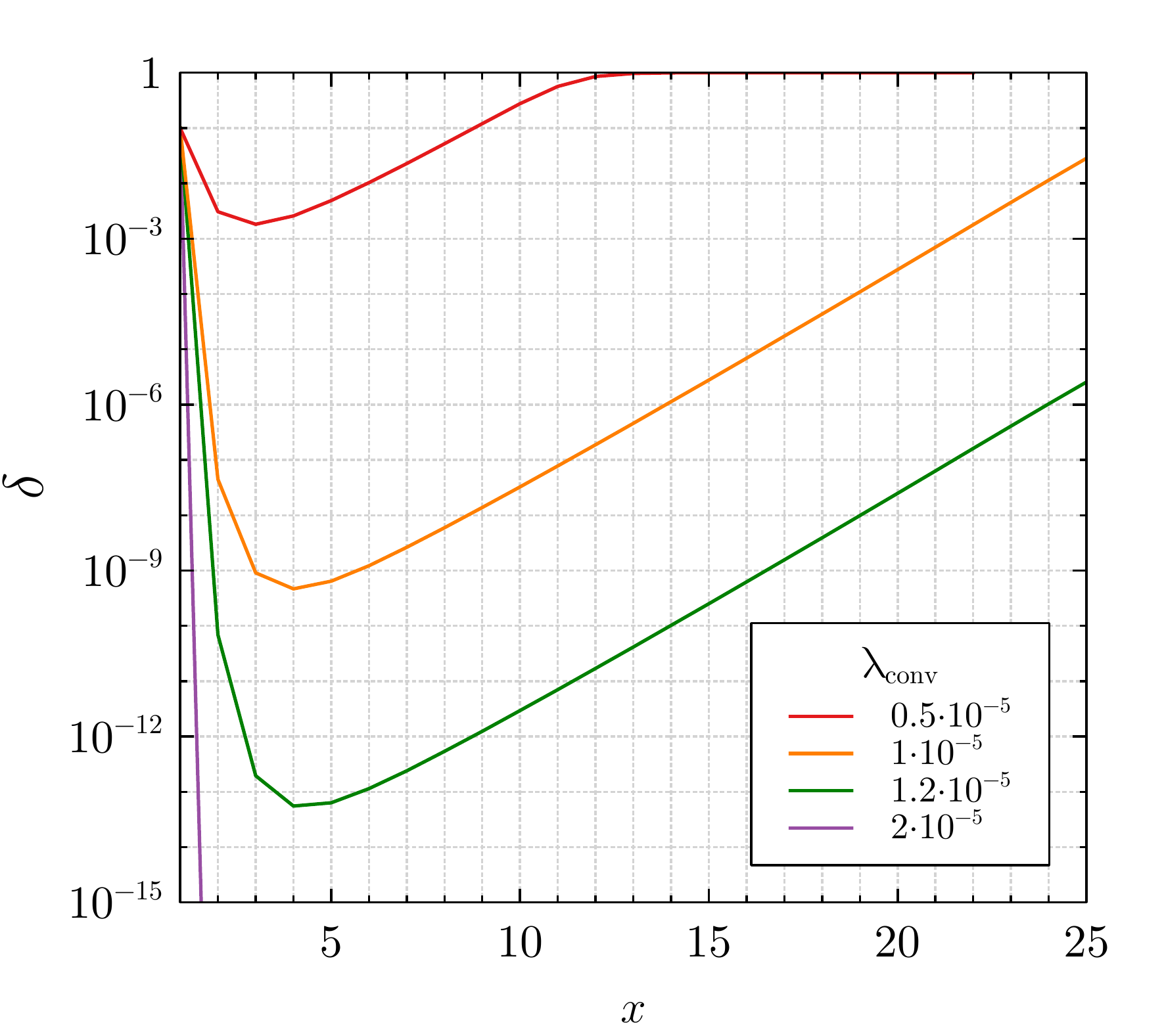}}
	\end{minipage}
	\begin{minipage}[h]{0.45\linewidth}
		\center{\includegraphics[width=1\textwidth]{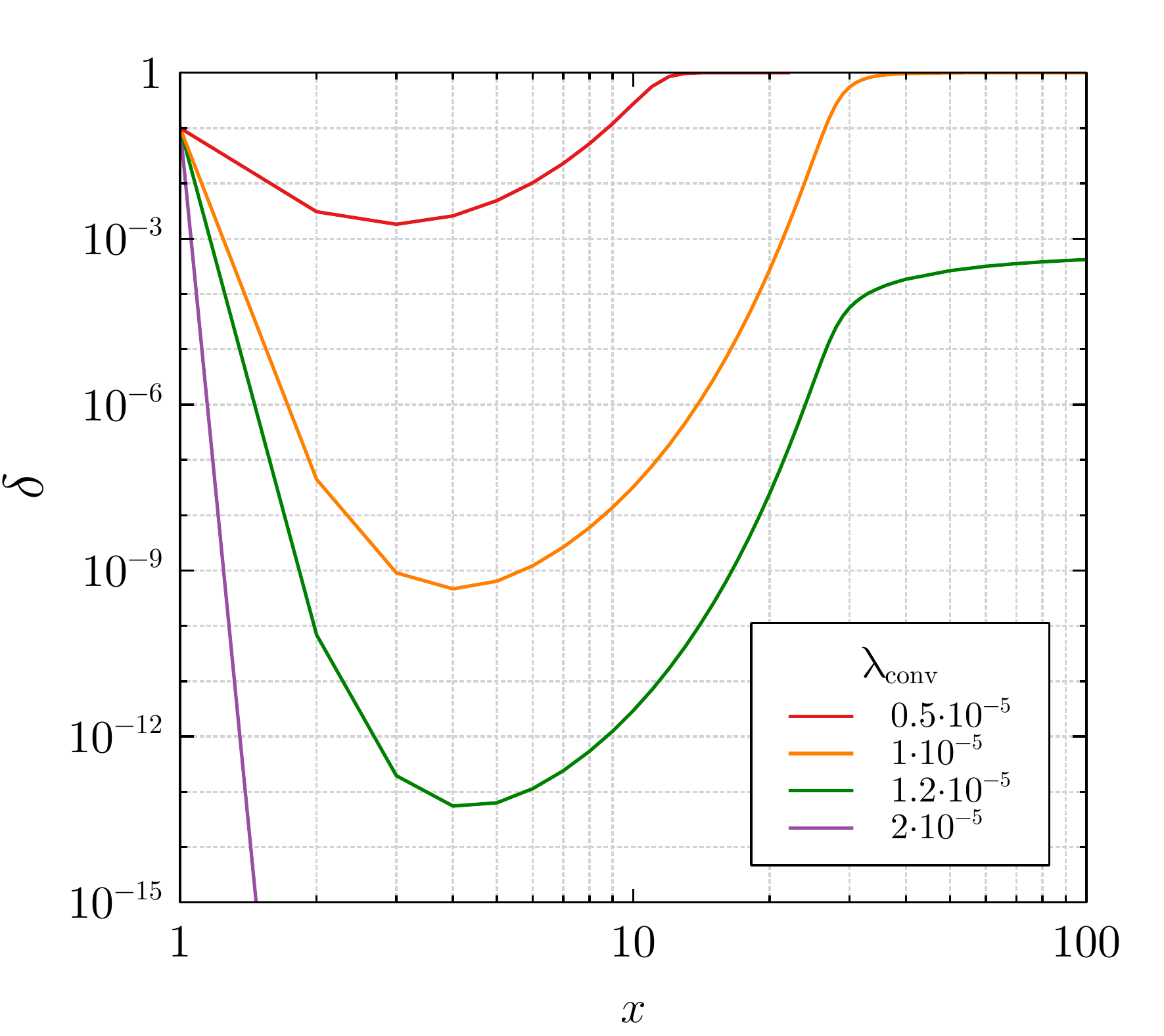}}
	\end{minipage}
	\begin{minipage}[h]{0.45\linewidth}
		\center{\includegraphics[width=1\textwidth]{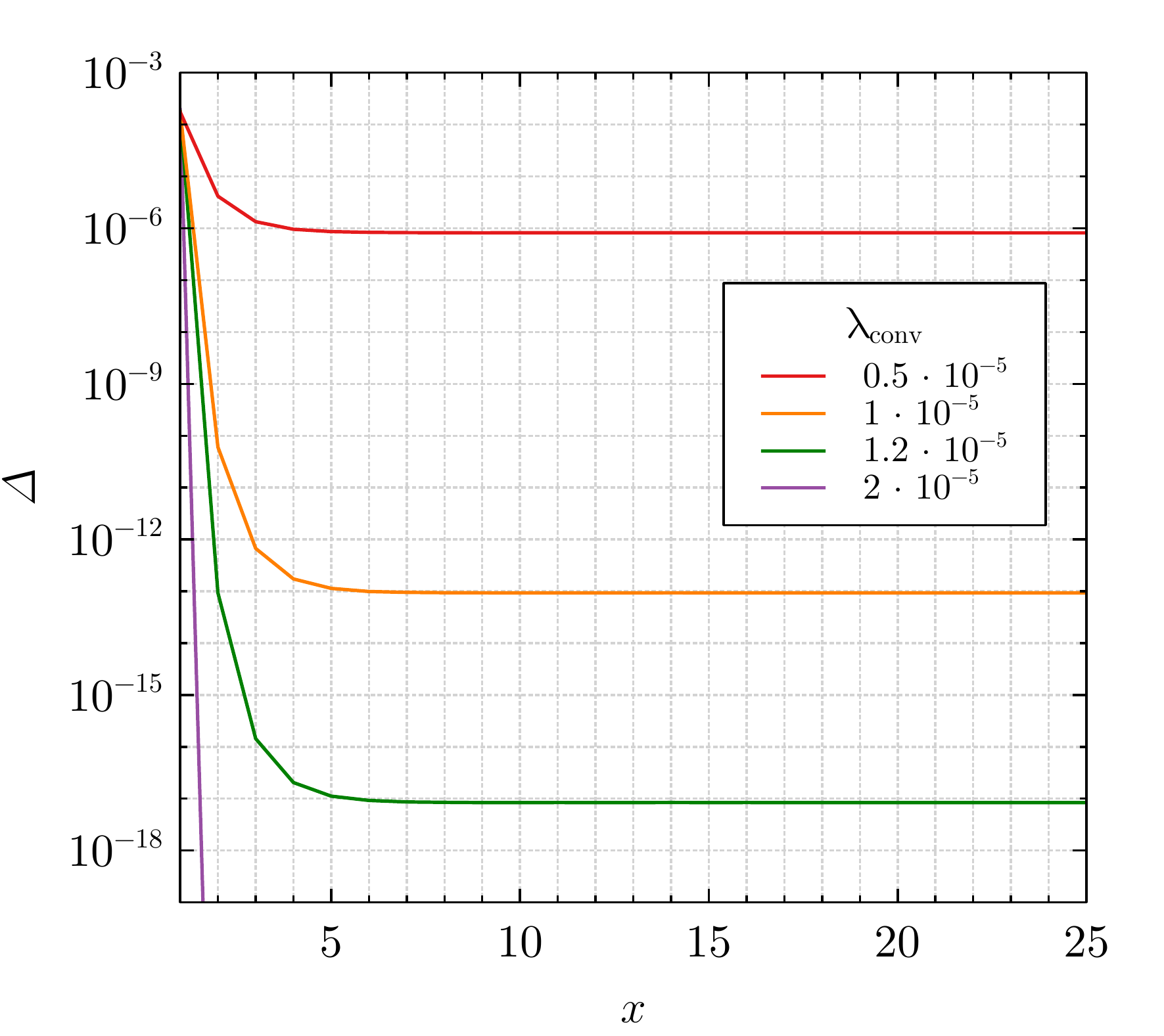}}
	\end{minipage}
	\begin{minipage}[h]{0.45\linewidth}
		\center{\includegraphics[width=1\textwidth]{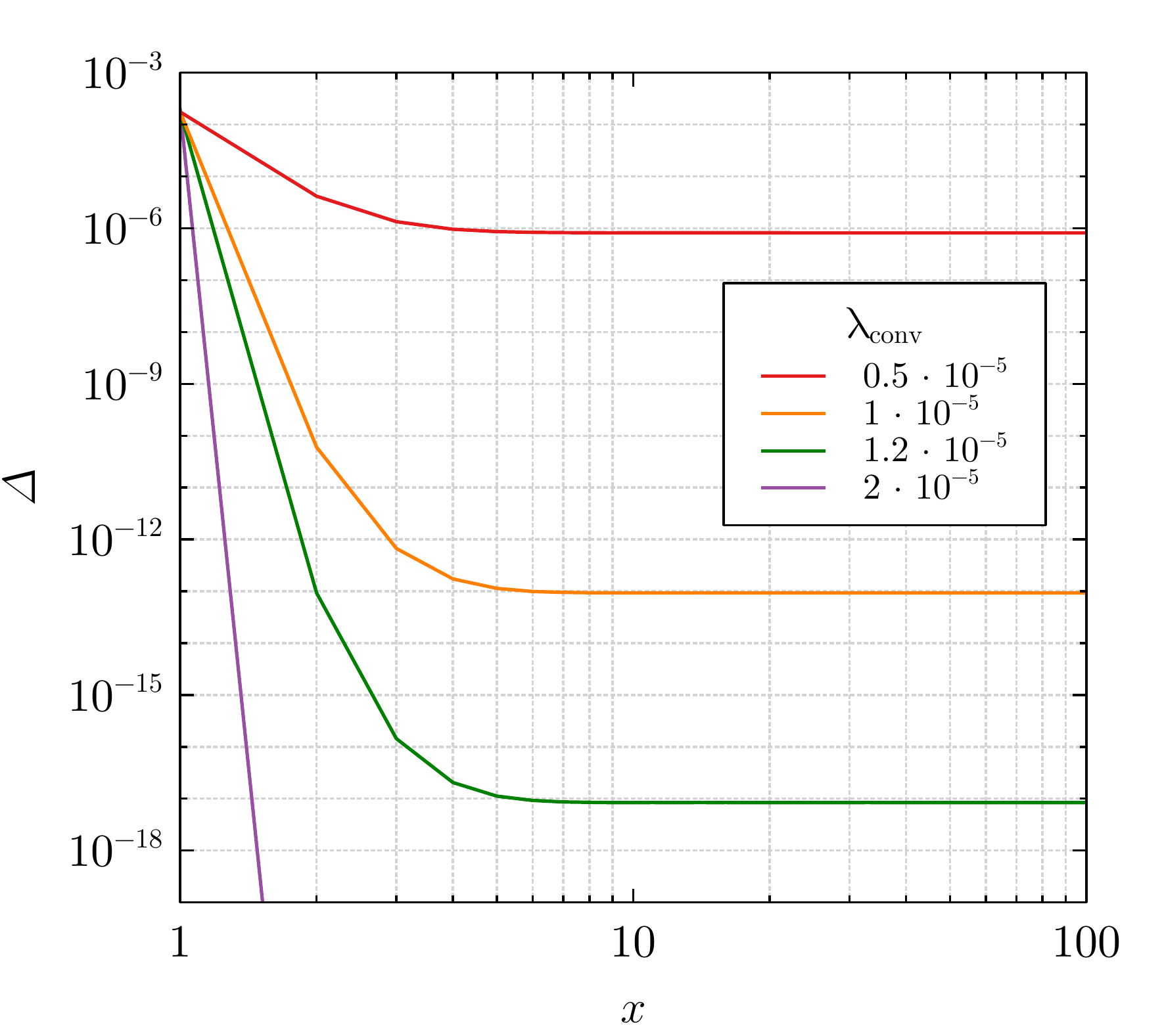}}
	\end{minipage}
	\caption{Top row: Asymmetry of DM $\delta$ as a function of $x$ for different values of the conversion constant $\lambda_{\rm conv}$ in linear scale (\textit{left}) and logarithmic scale (\textit{right}). The other initial conditions have their default values. Bottom row: the same for the absolute value of the asymmetry $\Delta$.
}
	\label{plot_lambda} 
\end{figure}

Fig.~\ref{plot_lambda} shows the main feature we discussed above: the $x$ evolution
of the absolute asymmetry $\Delta$ and the relative asymmetry $\delta$ as well as the modifications induced by the conversion process $\varphi\varphi \leftrightarrow \bar\varphi\bar\varphi$.
One sees that when $\lambda_{\rm conv}$ is sufficiently small (red curve), the asymmetry first drops but then quickly restores to unity,
and the relic density freezes out at values of $x$ smaller than the typical symmetric DM value $x \approx 25$.
For $\lambda_{\rm conv} \approx 10^{-5}$ (yellow curve), the minimum gets deeper, nevertheless the asymmetry fully recovers
at later stages, around $x \approx 30$.
This is the boundary value at which the recovery completes just in time when the total density freeze out takes place.

For larger values of the conversion coupling (green curve), depletion of the asymmetry is more intense and extends to slightly larger values of $x$.
The recovery stage starts with a much lower value of $\delta$ and it does not reach unity by the moment, when the total density freezes out.
There is some residual growth of $\delta$ in the flattening stage, but the final value of $\delta(\infty) \ll 1$.
Taking $\lambda_{\rm conv}$ to be twice larger than the boundary value (magenta curve) induces conversion effects
so strong that they irreparably wash out the asymmetry. Also note that the left part of Fig.~\ref{plot_lambda} shows very nice correspondence between the numerical results and analytical predictions for the recovery stage, described by Eq.~\eqref{delta_restore}.

\begin{figure}[h!]
\center{
\includegraphics[width=0.44\textwidth]{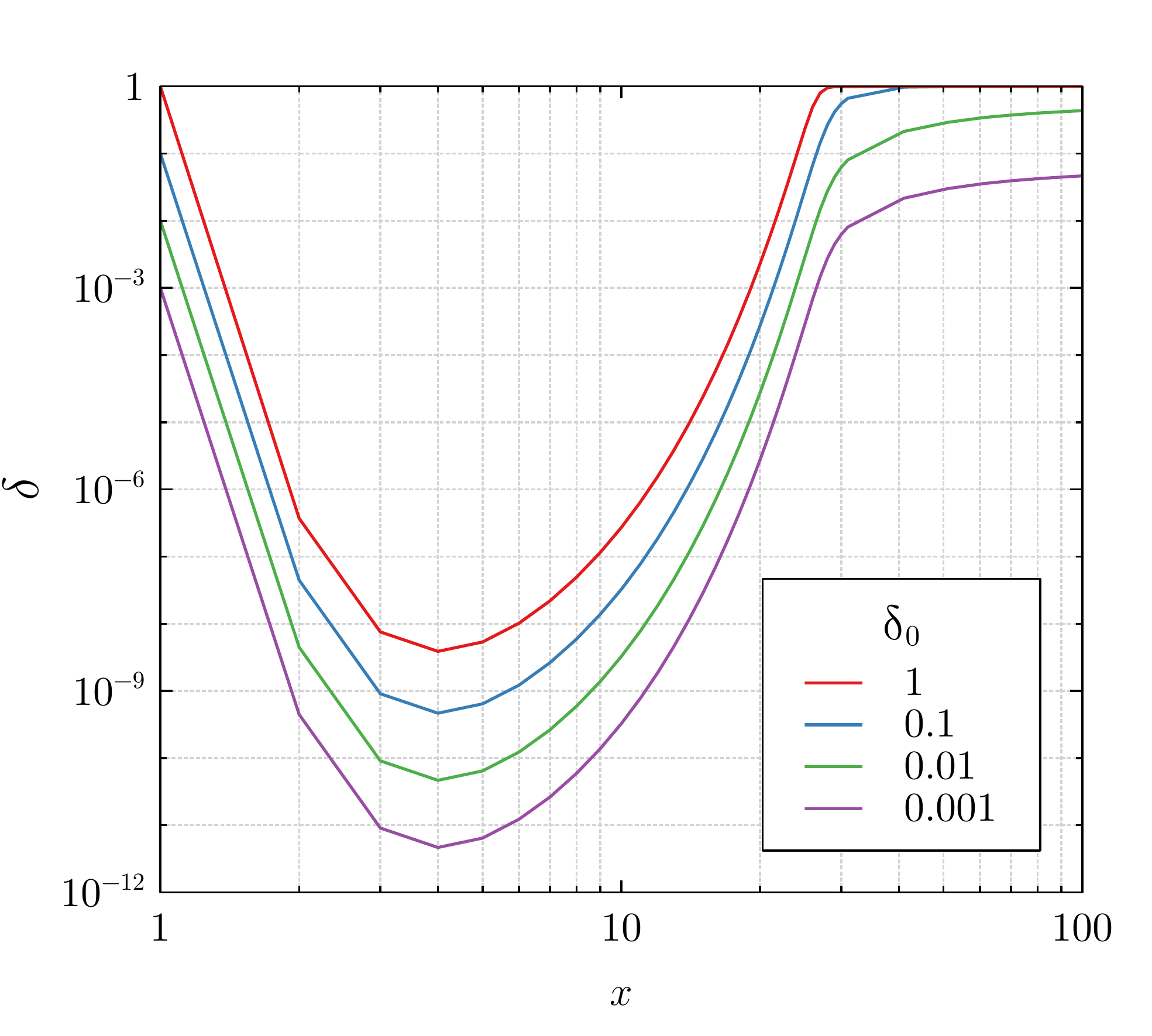}
\includegraphics[width=0.55\textwidth]{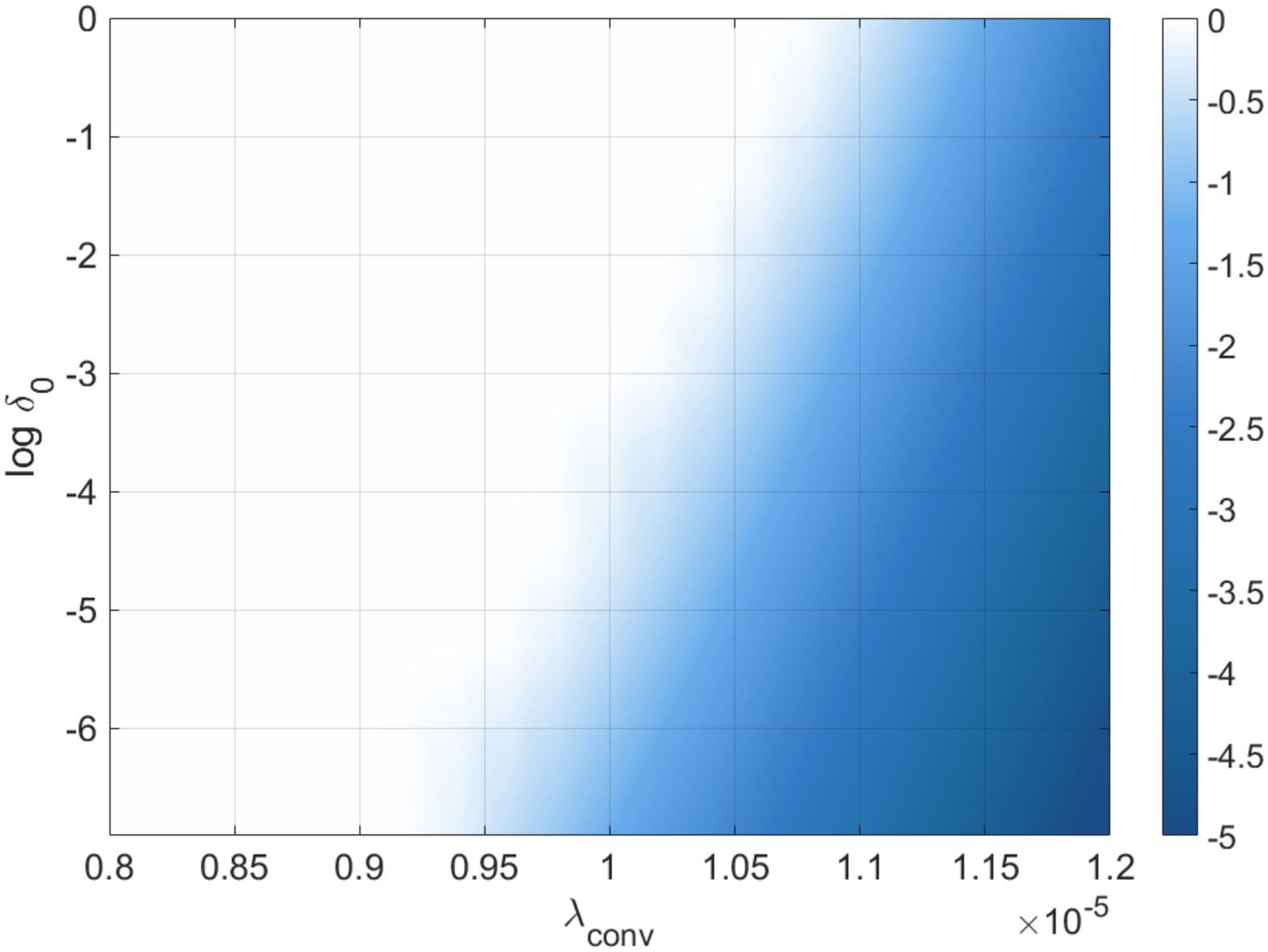}}
	\caption{Left: Asymmetry of DM $\delta$ as a function of $x$ for different values of the initial asymmetry $\delta_0$. The other initial conditions have their default values. The value of the conversion constant is $\lambda_{\rm conv} = 10^{-5}$.
	Right: The logarithm of the final asymmetry $\log \delta \equiv \log_{10} \delta$ (the colorbar) as a function of the conversion constant $\lambda_{\rm conv}$ and the logarithm of the initial asymmetry $\log \delta_0$. The other initial conditions have their default values. }
	\label{plot_delta_in} 
\end{figure}

In Fig.~\ref{plot_delta_in}, left, we show the role of the starting value of the asymmetry $\delta_0$ at $x_0 = 1$ for the fixed conversion
strength $\lambda_{\rm conv} = 10^{-5}$.
One can see the neat proportionality of the curves, which confirms the results of the qualitative analysis:
during each stage, the value of asymmetry $\delta$ drops or increases by certain factor, unless it approaches unity.
This proportionality is driven essentially by the fact that $d\delta/dx$ is proportional to $\delta$ itself.

These three plots, Figs.~\ref{plot_lambda} and \ref{plot_delta_in}, illustrate our main conclusion:
in order for the asymmetry to get restored close to 100\%, one needs to select not too small $\delta_0$
and not too strong $\lambda_{\rm conv}$.
Both of these values are the free parameters of the model, and 
Fig.~\ref{plot_delta_in}, right, demonstrates the boundary between the two regimes on their plane.

The exact boundary, of course, depends on the moment $x_0$ at which certain initial asymmetry $\delta_0$ is postulated.
In all plots so far we assumed that $x_0 =1$. If the asymmetry is generated at smaller temperatures (larger values of $x_0$),
then the asymmetry can get fully restored even in the presence of more intense conversion process.
This dependence is rather sharp and can be observed already in Fig.~\ref{plot_delta_in}, left.
If one starts at $x_0=1$ with $\delta_0 = 10^{-3}$ (magenta curve), then $|\lambda_{\rm conv}| = 10^{-5}$
keeps the final $\delta$ at the few percent level.
However, if one postulates the same $\delta_0 = 10^{-3}$ at slightly lower temperature, with $x_0=1.5$,
then the final asymmetry easily recovers to nearly unity.
This strong sensitivity to the initial conditions is unavoidable when the depletion and recovery factors
at the intermediate stages display exponential behavior.

\section{Indirect detection}
\label{indirect}

In the classical symmetric DM scenario, the single process --- DM annihilation into the SM fields --- determines, albeit in different kinematics, both the relic abundance and the possible indirect detection (ID) signal, 
which may be observable today. 
In the asymmetric DM scenarios, this relation is looser. In particular, in the highly asymmetric case of CP4 3HDM,
even if there remains a vanishingly small amount of $\bar\varphi$ after freeze-out, it can be regenerated 
later via $\varphi\varphi\to \bar\varphi\bar\varphi$ conversion, and only then annihilate to the SM fields.
In this Section, we investigate whether this two-stage process can produce an observable ID signal.

Let us assume that the relic abundance of $\bar\varphi$ is much smaller than the abundance of $\varphi$.
Then, in the cosmologically modern epoch, the evolution of the local density of $\bar{\varphi}$ is given by the following equation
\be
\fr{\partial \bar{n}(t)}{\partial t} = \svC n^2 - \svSM \, n\bar{n}(t) \, .
\label{phi_bar_evol}
\ee
It is implicitly assumed that all the quantities in Eq.~\eqref{phi_bar_evol} 
including the velocity-averaged cross sections can be spatially dependent. 
The density of $\varphi$ can be considered as approximately constant in time.
Indeed, even if we consider symmetric DM in very dense regions, such as DM cusps at the centers of galaxies, where the density can be up to $\rho \sim 10^3 \GeV/\cm^{3}$ (see e.g. Fig.~1 in \cite{Cirelli:2010xx}), and take the maximal value of $\svSM \sim 10^{-20} \cm^3/\s$ allowed by the unitarity bound for $m \sim 100 \GeV$, the characteristic timescale of annihilation $\tau = \left(n\svSM\right)^{-1}$ would be around $10^3$ Gyrs, which already exceeds the lifetime of the Universe by two orders of magnitude. 
Our aim is to understand whether this large and essentially constant amount of DM 
can lead to efficient conversion and observable ID signal.

Taking into account our assumption, we can easily solve Eq.~\eqref{phi_bar_evol} for $\bar{n}$:
\be
\bar{n}(t) = \bar{n}_0 \exp\left(-t/\tau \right) + n\fr{\svC}{\svSM} \left[ 1 - \exp\left(-t/\tau\right) \right] \, .
\ee
Here, $\bar{n}_0$ is the remaining density of antiparticles after the freeze out, 
which can be extremely low as it is exponentially sensitive to the cross section of annihilation into SM, Eq.~\eqref{final-r}.
The time $t$ is of the order of the Universe's lifetime, $t \approx 10$ Gyrs, and it is much smaller that $\tau$.
This allows us to simplify the solution for $r \equiv \bar n/ n$ to
\be
r(t) \approx \, r_0  +  \left(\fr{\svC}{\svSM} - r_0\right) {t \over \tau} \, .
\label{anti-density}
\ee
Notice that $\svC$ is calculated here in the present cosmological epoch. Due to its proportionality to 
the relative velocity, it is roughly two to three orders of magnitude smaller than it was at the freeze-out time
(the exact value would be sensitive to the details of the DM velocity distribution in galaxies).

The bracket here can have either sign, depending on $r_0$, 
the relative amount of $\bar\varphi$ after the freeze out, and on the ratio of the two cross sections.
Consider first the case when the sign is negative, which can happen if $r_0$ is not negligibly small.
It corresponds to the evolution scenario where the relative amount of $\bar\varphi$ was initially fixed at the level
of Eq.~\eqref{final-r} after the freeze-out process, but then, in the structure formation epoch,
the annihilation gradually switched back on, leading to further depletion of $\bar\varphi$'s and producing ID signal.
In this regime, the amount of $\bar\varphi$ does not change much during the lifetime of the Universe: 
$r(t) \approx r_0$, and the conversion process does not play any role in forming the ID signal.
The annihilation rate per unit volume is given by 
\be
\gamma_{\rm asym} = n\bar{n} \svSM \, .
\ee
We now compare this with the rate of annihilation in the conventional symmetric case, which we write as $\gamma_{\rm sym} = n^2 \svSMsym$, 
with $\svSMsym \approx 10^{-26} \cm^3/\s$ being the commonly adopted value of the velocity-averaged cross section that yields the correct value of the DM relic abundance. Here we would like to stress again that the relic abundance in the asymmetric scenario does not necessarily depend on $\svSM$. 
Thus the value of $\svSM$ can be larger than $\svSMsym$ mentioned above and still be compatible with the observed relic abundance.
Neglecting the possible difference in the spatial distribution of the annihilation signal 
that could appear in the presence of conversion, we get a simple expression
\be
\frac{\gamma_{\rm asym}}{\gamma_{\rm sym}} = r_0 \, \left(\frac{\svSM}{10^{-26} \cm^3/\s}\right)\,.
\ee
Can this ratio be larger than 1? Naively, one may think of the situation when $r_0$ is only moderately small,
while the SM annihilation cross section is much higher in our case than in the symmetric DM scenario.
Unfortunately, this situation is not possible. The value of $r_0$, which is the same as $r(\infty)$ in Eq.~\eqref{final-r},
does depend on the SM annihilation (parameter $a$) in a very essential way.
By increasing $\svSM$ with respect to $\svSMsym$, one would strongly suppress $r_0$,
deteriorating the ID signal.

One can also consider a different regime, when $r_0$ is extremely small due to 
a rather large annihilation cross section. Then, the sign of the bracket in Eq.~\eqref{anti-density} is positive.
This is a novel scenario, in which particles $\bar\varphi$, whose amount was vanishingly small after the freeze-out,
are being regenerated in the present cosmological epoch.
To consider the extreme case, we set $r_0 \to 0$ and obtain

\be
\frac{\gamma_{\rm asym}}{\gamma_{\rm sym}} = n \svC t \, \left(\frac{\svSM}{10^{-26} \cm^3/\s} \right)\, .
\ee
In this regime, the ID signal is time-dependent; it is predicted to be stronger now 
than during the structure formation epoch. However, its absolute value is even lower than in the previous case.
An order-of-magnitude estimate for the values mentioned above gives
\be
\fr{\gamma_{\rm asym}}{\gamma_{\rm sym}} \sim 10^{-21} \left(\fr{\rho}{10^3 \GeV/\cm^{3}}\right) \left(\fr{100 \GeV}{M}\right) \left(\fr{\svC}{10^{-34} \cm^3/\s}\right) \left(\frac{\svSM}{10^{-26} \cm^3/\s} \right)\, .
\label{rate_compar}
\ee
The bottom line is that, for a very broad range of velocity-averaged conversion cross sections, 
the signal from dense DM regions in the asymmetric case is suppressed by many orders of magnitude 
compared to the symmetric DM scenario. 
Due to the strong constraints on the efficiency of the conversion process, we simply cannot generate 
any sufficient amount of $\bar\varphi$ in the given time.

\section{Conclusions}

Dark matter can be stabilized by a $CP$ symmetry --- provided it is a symmetry of an order higher than 2
and it is preserved by the vacuum.
CP4 3HDM, the three-Higgs-doublet model featuring $CP$ symmetry of order 4, is the simplest
framework producing such DM candidates \cite{Ivanov:2015mwl}.
Motivated by this form of stabilization, which was not observed in beyond the SM models before,
we investigate its consequences for the DM evolution and freeze-out process.

Due to the unavoidable mass degeneracy \cite{Haber:2018iwr}, the DM sector contains
two stable complex fields $\varphi$ and $\bar\varphi$,
which are CP4 eigenstates but which are not particle and antiparticle of each other.
It is a key consequence of $CP$ conservation of this model that $\varphi\bar\varphi$
can annihilate into SM fields, while $\varphi\varphi$ and $\bar\varphi\bar\varphi$ cannot.
However, these two pairs can turn into each other, $\varphi\varphi \leftrightarrow\bar\varphi\bar\varphi$,
and this conversion process can strongly affect the DM evolution.
It offers a rare example of a DM model in which self-interaction in the dark sector can significantly affect the cosmological observables
and, therefore, can be constrained.

These considerations led us to investigate the {\em asymmetric} DM regime, in which some asymmetry between 
$\varphi$ and $\bar\varphi$ is generated at $T \sim m$ by a mechanism external to CP4 3HDM.
The presence of conversion process makes this model clearly different 
from the more conventional asymmetric DM models \cite{Graesser:2011wi,Iminniyaz:2011yp,Zurek:2013wia,Petraki:2013wwa}. 
The questions we wanted to answer included: does the asymmetry survive in the presence of conversion?
Is the final asymmetry large or small? What is the role of the initial conditions and of the physical processes?
In the strongly asymmetric case, can this model produce an observable indirect detection signal
through the two-stage conversion-then-annihilation process?

To answer them, we presented a detailed analytical and numerical analysis of the system of coupled Boltzmann equations,
without and with conversion. For the numerical analysis, we wrote our own code 
since the existing packages are incapable of dealing with this model.
We found that the relative asymmetry $\delta$ first sharply drops from its initial value, driven by the conversion process.
When conversion effectively switches off, the relative asymmetry enters the recovery stage,
and finally freezes out, either at some small value $\delta(\infty) \ll 1$, implying $n(\bar\varphi) \approx n(\varphi)$, 
or very close to 1, which means $r = n(\bar\varphi)/n(\varphi) \ll 1$.
We established the boundary between the two regimes and showed how it depends on the initial conditions
and on the intensity of the conversion process.
In particular, we established that, in order for the asymmetry to recover to sizable values, 
the conversion cross section must be several orders of magnitude smaller than the annihilation one.

We also checked the intensity of the ID signal in both regimes and found that the expected signal
is much smaller than in the symmetric DM scenario. Although it does not offer a novel observational signature
of this model, it may help to resurrect scenarios in which one would normally expect a very strong ID signal 
in the usual symmetric DM case.

We stress that we did not aim to perform a systematic scan of the parameter space of CP4 3HDM.
Nor do we claim that this model is superior to others in resolving some long-standing DM-related issues, such as small-scale problems.
Rather, we wanted to investigate the {\em phenomenon} itself. We believe that qualitative features 
of the thermal DM history found in this model can arise in other models with the conversion processes
competing with annihilation, which deserves further investigation.

\bigskip
I.P.I. acknowledges funding from the the Portuguese
\textit{Fun\-da\-\c{c}\~{a}o para a Ci\^{e}ncia e a Tecnologia} (FCT) through the FCT Investigator 
contract IF/00989/2014/CP1214/CT0004 under the IF2014 Programme,
and through the contracts UID/FIS/00777/2013, UID/FIS/00777/2019, CERN/FIS-PAR/0004/2017 and PTDC/FIS-PAR/29436/2017,
which are partially funded through POCTI (FEDER), COMPETE, QREN, and the EU.
I.P.I. also acknowledges support from National Science Center, Poland, via the project Harmonia (UMO-2015/18/M/ST2/00518).
I.P.I. thanks University of Liege for hospitality and stimulating atmosphere. M.L. thanks A. Bhattacharya for useful advice on C++.
The work of M.L. is supported by a FRIA grant (F.N.R.S.).

\appendix

\section{DM states and interactions in CP4 3HDM}\label{appendix-CP4-3HDM}

The two lightest neutral fields $h$ and $a$ in Eq.~\eqref{fields-basis} 
are the DM candidates protected by CP4 symmetry. In principle, all interactions can be written in terms
of these real fields. However, they are not CP4-eigenstates, as they transform under CP4 as $h \toCP -a$ and $a\toCP h$. 
Since CP4 is conserved, we combine them into CP4 eigenstates \cite{Ivanov:2015mwl,Aranda:2016qmp}:
\be
\varphi = {1\over \sqrt{2}}(h + i a)\,, \quad \varphi \toCP i\varphi\,.\label{J-eigenstates1}
\ee
We stress that this peculiar form of CP transformation does not involve 
any redefinition of the CP symmetry {\em per se};
it is the same CP4 as in Eq.~(\ref{CP4}), just applied to the physical states after minimization \cite{Aranda:2016qmp}.

Much care must be taken when interpreting particle/antiparticle assignments
of the corresponding one-particle states.
Express the complex field $\varphi$ via creation and annihilation operators,
\be
\varphi(\vec{x},t) = \int \tilde{dp} \left[a(\vec{p}) e^{-ipx} + b^\dagger(\vec{p})e^{ipx}\right]\,,\label{phi}
\ee
where $px \equiv E t - \vec{p} \vec{x}$ and $\tilde{dp} \equiv d^3 p/[2E(2\pi)^3]$.
The standard assignment of the $C$-transformation, which would map a complex field to its conjugate,
would interchange operators $a$ and $b$ and would make 
the one-particle states $a^\dagger|0\rangle$ and $b^\dagger|0\rangle$ antiparticles of each other.
In our case, the definition of the $CP$-symmetry via Eq.~(\ref{J-eigenstates1}) 
forces us to accept that the one-particle state $a^\dagger|0\rangle$,
which we denote as $\varphi$ in the main text, is its own antiparticle.
The other one-particle state with the same mass, $b^\dagger|0\rangle$, which we denote as $\bar{\varphi}$,
is a {\em different} state and is also its own antiparticle, see discussions in \cite{Aranda:2016qmp,Haber:2018iwr}. 
We stress again that we abused the language in the main text of this paper and 
called $\bar\varphi$ ``antiparticles''. This was done only to facilitate wording;
in reality, $\varphi$ and $\bar\varphi$ are not antiparticles of each other,
but are two distinct mass-degenerate particles.

The full Lagrangian can be rewritten in terms of these complex fields 
rather than $h$ and $a$ \cite{Kopke:2018vyw,Haber:2018iwr}. 
In this paper, we do not perform a detailed scan of the parameter space of the model
but rather focus on a particular phenomenon.
Thus, we find it sufficient to consider the DM mass above 100 GeV and to consider, 
for annihilation, only the channels $\varphi\bar{\varphi} \to WW$, $ZZ$.
Their diagrams include the pointlike interactions and the $s$-channel
annihilation via the SM-like Higgs boson coming from
\be
{\cal L} \supset \left({g^2 \over 2} W_\mu^+W_\mu^- + {\bar g^2 \over 4} Z_\mu Z^\mu\right) \varphi^*\varphi 
+ \left(\varv h_{125} + {1 \over 2}h_{125}^2\right) \lambda_{346}\varphi^* \varphi\label{interaction-1}\,,
\ee
where $\bar{g}^2 \equiv g^2 + g^{\prime 2}$, with $g$, $g'$ being the gauge coupling constants,
as well as the $t$ and $u$-channel annihilation diagrams which involve triple interactions 
linking $\varphi$'s with the gauge bosons and the heavier inert scalars, either neutral or charged. 
In numerical calculations, we used all these diagrams. To show an estimate
of the cross section we expect at small incoming velocities and large $m \gg m_W$, 
one can neglect the $t$ and $u$ channel diagrams and express the cross section of the DM annihilation 
into the SM fields as
\be
\sigma_{\mathrm SM} \approx {1 \over 32 \pi m^2} {1\over v} \left({g^4 \over 4} + {\bar {g}^2\over 8} + {3\over 2}\lambda_3^2\right)\,.\label{annihilation-estimate}
\ee
Here, the first two contributions in the brackets come from the transverse polarizations of the $W$ and $Z$ pairs;
they combine to $\approx 0.08$.
The last contribution is the large-$m$ estimate of the cross section of the longitudinal 
gauge boson pairs.\footnote{For a pedagogical exposition, with the example of the inert doublet model, 
	of how $\lambda_3$ arises in the annihilation amplitude, see Section~3.2 of Ref.~\cite{Hambye:2009pw}}
Thus, for $\lambda_3 = 0.9$ and $m=250$ GeV used in numerical calculations, we get
\be
\sigma_{\mathrm SM}v \approx 1.6 \cdot 10^{-7}\ \mbox{GeV}^{-2} \approx 2\cdot 10^{-24} \cm^3/\s\,.
\ee
Of course, in the funnel region, $m \approx m_{h_{125}}/2$, the annihilation cross section will be resonantly enhanced,
while at even smaller masses it will be strongly suppressed.

For the conversion process, we first write the self-interaction potential for $\varphi$:
\be
V_{si}= {\lambda_2 \over 8}(\varphi^4 + \varphi^{* 4} + 6 \varphi^2 \varphi^{* 2}) 
- {\lambda_3' + \lambda_4' - 2 \Re \lambda_8\over 16} (\varphi^2 - \varphi^{*2})^2  
+ i {\Im \lambda_9 \over 4} (\varphi^4 - \varphi^{*4})\,.\label{self-phiphi}
\ee
The $\varphi^4$ interaction terms lead to the following amplitude of conversion process 
$\varphi\varphi \to \bar\varphi\bar\varphi$:
\be
{\cal M}_{\rm conv} \equiv -i \lambda_{\rm conv} = -i{3 \over 2} \left(2\lambda_2 - \lambda_3' - \lambda_4' + 2\Re \lambda_8
+ 4i \, \Im \Lambda_9\right)\,,
\ee
whose cross section is
\be
\sigma_{\rm conv} = \fr{|\lambda_{\rm conv}|^2}{128 \pi m^2}\,.\label{sigma-conv}
\ee 

\bibliographystyle{JHEP}
\bibliography{Asymmetry}

\end{document}